\documentclass[twocolumn,showpacs,preprintnumbers,amsmath,amssymb,superscriptaddress,nofootinbib,groupedaddress,floatfix]{revtex4-1}
\usepackage{graphicx} 
\usepackage{dcolumn} 
\usepackage{bm} 
\usepackage{amssymb} 
\usepackage{amsmath} 
\usepackage{amsfonts}
\usepackage{color}
\usepackage{comment}
\usepackage{rotating}
\usepackage{multirow}
\usepackage{ctable}
\usepackage{colortbl}
\usepackage{afterpage}

\usepackage{array}
\newcommand{\PreserveBackslash}[1]{\let\temp=\\#1\let\\=\temp}
\newcolumntype{C}[1]{>{\PreserveBackslash\centering}p{#1}}
\newcolumntype{R}[1]{>{\PreserveBackslash\raggedleft}p{#1}}
\newcolumntype{L}[1]{>{\PreserveBackslash\raggedright}p{#1}}

\definecolor{darkgreen}{RGB}{50,150,0}

\newcommand{\mnras}{Mon. Not. R. Astron. Soc.}

\begin{document}

\title{$H_0$ Tension, Swampland Conjectures
\\and the \\Epoch of Fading Dark Matter}
\author{Prateek Agrawal, Georges Obied, Cumrun Vafa}
\affiliation{Jefferson Physical Laboratory, Harvard University, Cambridge, MA 02138, USA}

\begin{abstract}

Motivated by the swampland dS conjecture, we consider a rolling scalar field as the source of dark energy.
Furthermore, the swampland distance conjecture suggests that the rolling field will lead at late times to an exponentially light tower of states.
Identifying this tower as residing in the dark sector suggests a natural coupling of the scalar field to the dark matter, leading to a continually reducing dark matter mass
as the scalar field rolls in the recent cosmological epoch.
The exponent in the distance conjecture, $\tilde{c}$, is expected to be an $\mathcal{O}(1)$ number.
Interestingly, when we include the local measurement of $H_0$, our model prefers a non-zero value of the coupling $\tilde{c}$ with a significance of $2.8\sigma$ and a best-fit at $\tilde{c} \sim 0.3$.
Modifying the recent evolution of the universe in this way improves the fit to data at the $2\sigma$ level compared to $\Lambda$CDM. This string-inspired model automatically reduces cosmological tensions in the $H_0$ measurement as well as $\sigma_8$.

\end{abstract}

\maketitle

\section{Introduction}
\label{sec:int}

Recently the swampland program \cite{Vafa:2005ui} has been studied rather intensively from various viewpoints (see the review articles \cite{Brennan:2017rbf,Palti:2019pca}).  In particular a conjecture has been advanced about meta-stable de Sitter as belonging to the swampland~\cite{Obied:2018sgi}. Namely for scalar field potentials arising from a consistent quantum gravity, we have a bound\footnote{There has been refined versions of this conjecture; see in particular~\cite{Garg:2018reu,Ooguri:2018wrx}. In this paper we explore a late time cosmology satisfying the slope condition on the potential but it is conceivable that this is based on satisfying the second clause of the refined de Sitter conjecture as in CITE. }
$$|\nabla V|\geq c \ V.$$
where $c\sim \mathcal{O}(1)$ in Planck units.
This conjecture motivated alternative descriptions of the present epoch of cosmology~\cite{Agrawal:2018own} as being realized through a rolling scalar field, which is sometimes called the quintessence field. Studying quintessence models is natural, independently of the validity of the dS swampland conjectures. Here we would like to study quintessence models in view of yet another swampland conjecture, the distance conjecture~\cite{Ooguri:2006in}, which states that if a scalar field moves a distance $\Delta \phi \geq \mathcal{O}(1)$ in Planck units, a tower of light states emerges (see related work~\cite{Grimm:2018ohb,Heidenreich:2018kpg,Blumenhagen:2018hsh}).  Namely masses in the tower scale as the field rolls for large values of $\phi$ by
$$m_i(\phi)\sim m_i(0)\,\exp(-{\tilde c}\,\phi)$$
where ${\tilde c}\sim \mathcal{O}(1)$ in Planck units.

Aspects of this conjecture were already studied in this context in~\cite{Agrawal:2018own}, where it was pointed out that in this scenario our universe will undergo a transition in a time scale of order the Hubble time in one of two possible ways: (i) either the dark energy becomes negative or (ii) the scalar field will roll a large distance which leads to the emergence of a light tower of states with which the quintessence field interacts strongly. As is well known the quintessence field cannot be interacting strongly with the visible sector.  Therefore this light tower of states must reside in the dark sector.  It is natural to ask whether we have already begun to experience the emergence of this light tower of states in the dark sector through the rolling of the quintessence field.
In this paper we aim to study this question.

On the other hand recent observations of local cosmology~\cite{riess2018new,Riess:2019cxk} have led to an unexpectedly large value of the present Hubble constant $H_0$ compared to that inferred from $\Lambda$CDM using the Cosmic Microwave Background (CMB) data~\cite{Ade:2015xua,Aghanim:2018eyx}. This so called Hubble tension is now significant at the $4.4\sigma$ level~\cite{Riess:2019cxk}. There has also been studies of the $H_0$ tension in the context of the dS swampland
conjecture~\cite{DiValentino:2017zyq,
Durrive:2018quo,
Yang:2018xah,
Tosone:2018qei,
Heisenberg:2018yae,
Akrami:2018ylq,
Raveri:2018ddi,
Elizalde:2018dvw,
Colgain:2018wgk,
Kaloper:2019lpl,
Colgain:2019joh}.
It turns out that quintessence models exacerbate this tension since the dark energy density decreases in recent times.  However if we assume that the tower of light states are already beginning to emerge in the dark sector, then as we will see, the story dramatically changes. Such a coupling leads to a reduction of mass, or fading, of dark matter which is compensated by a bigger value of dark energy.  The latter becomes  more noticeable in the present accelerating epoch, leading to an increase in $H_0$.  Moreover the fit to data improves by $2\sigma$ compared to $\Lambda$CDM. The $H_0$ value increases from $68.3$ in $\Lambda$CDM to $69.1$.  This improves the tension with the observed value of $H_0$ using supernovae data~\cite{riess2018new,Riess:2019cxk} but is not enough to resolve it. Further, the lower value of dark matter at late times leads to a lower value of $S_8$ which also ameliorates tension with weak lensing measurements.

What is remarkable is that introducing this ingredient, which is motivated from string theory, not only improves the fit to the data but also predicts a value for ${\tilde c}$ by best fit with experiments which is significantly different from $0$ but still order 1.  Namely we find the best-fit value
$$\tilde c \sim 0.3$$
for $c\sim \mathcal{O}(1)$.  This is probably the most important result of this paper: the quintessence field is best set as interacting with the dark sector in an exponential way with the value of the exponent close to 1.  This in turn leads to a more specific prediction about the future of our universe.

The organization of this paper is as follows: in section~\ref{sec:setup} we present the concrete setup of our model and present the results of our fit and the phenomenology in section~\ref{sec:results}. We conclude in section~\ref{sec:conclusion}.

\section{The Basic Setup}
\label{sec:setup}

\begin{figure}
    \centering
    \includegraphics[width=0.45\textwidth]{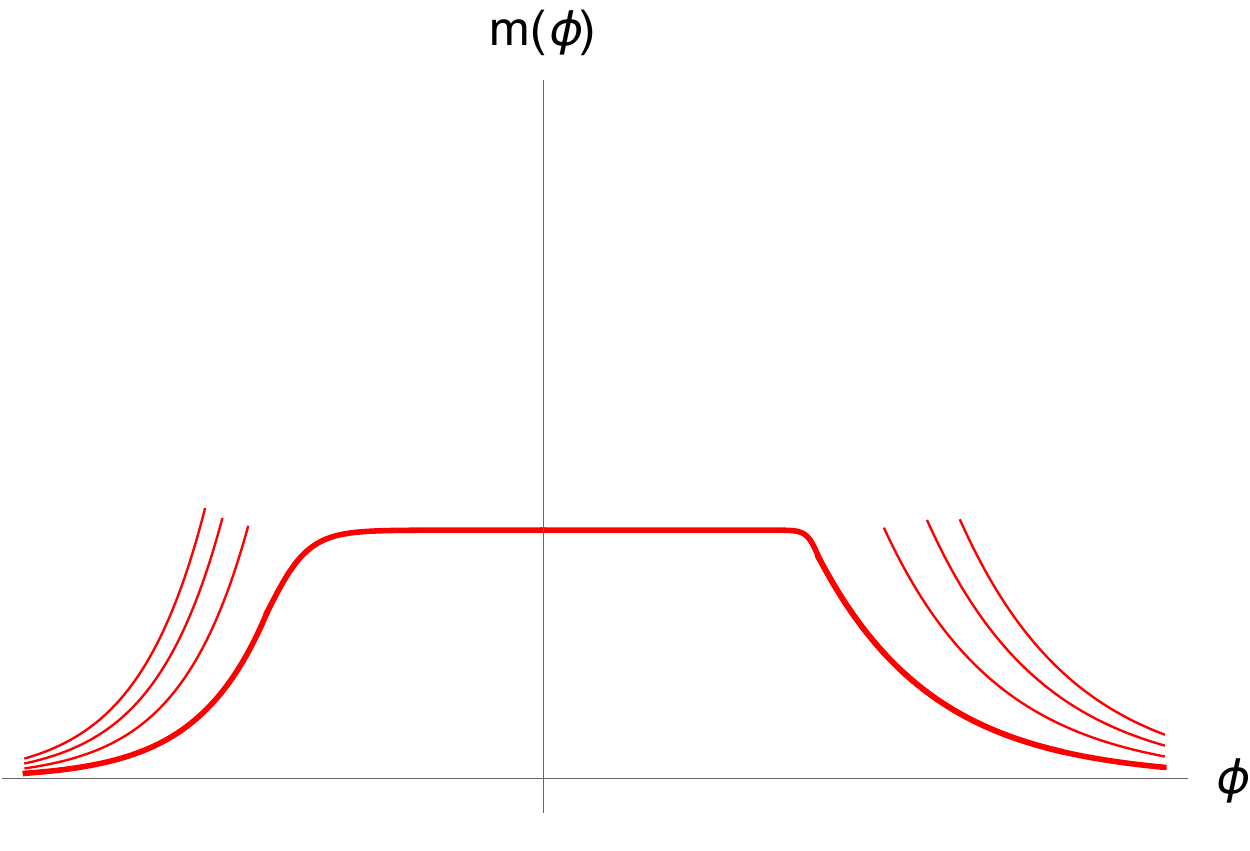}
    \caption{Schematic dependence of the mass gap $m(\phi)$ on the value of $\phi$. At large positive and negative field values the mass gap decreases exponentially, and a tower of states become light.}
    \label{fig:tower}
\end{figure}

Let us first briefly review the distance conjecture~\cite{Ooguri:2006in}. The distance conjecture is motivated by the observation that, when studying scalar fields in a large number of string theory vacua, light states emerge when the field travels distances $\phi \gg 1$ in Planck units.  In particular, a tower of light states emerges with a mass scale given by
$$m(\phi) \sim m_0\cdot\exp(-\tilde{c}\;\phi)$$
where $\phi$ labels the scalar field vev. This behavior is expected to hold when $\phi$ is sufficiently large.  Note on the other hand if we went to negative values of $\phi$ one may have expected the masses to increase, but the distance conjecture predicts that other light states will emerge whose mass again exponentially decreases as $\phi \ll -1$. Moreover for $|\phi|\lesssim \mathcal{O}(1)$ masses are not expected to be very sensitive to the field value $\phi$. See figure~\ref{fig:tower} for a depiction of expected behavior of $m(\phi)$. In fact a study of a class of models reveals exactly this behaviour (see figure 1 in~\cite{Baume:2016psm}). Therefore, if we take $m(\phi)$ to denote the mass gap as a function of $\phi$, a model that captures this behaviour for $\phi>0$ is
\begin{align}
\label{eq:mass}
m(\phi)
&=
m_0\left\{
\begin{array}{cc}
1 & 0 < \phi < \phi_0 \\
\exp[-\tilde c \ (\phi-\phi_0)] & \phi \geq \phi_0
\end{array}
\right.
\end{align}
which roughly models the behavior in figure~\ref{fig:tower}.
Alternatively, we can model the transition by using a smoother interpolating factor such as $ [1+\tanh( \gamma (\phi-\phi_0))]/2$
with $\gamma \gg 1$.

As already mentioned, the dS swampland conjecture suggests that dark energy is realized through a rolling scalar field $\phi$.  This was shown to be compatible with observation~\cite{Agrawal:2018own,Heisenberg:2018yae,Raveri:2018ddi} as long as $c \lesssim 0.6$. Moreover as was found in~\cite{Agrawal:2018own} the variation of $\phi$ until the present epoch is $\mathcal{O}(1)$.  Therefore it is natural to expect that we are entering a regime where a light tower of states associated with the dark sector is emerging. This motivates a model of dark matter as consisting of a particle of mass $m$
whose value changes as $m(\phi)$ given in Eq.~\ref{eq:mass} when $\phi$ rolls.  In other words the energy density of dark matter goes as
\begin{equation*}
\rho_{\rm DM} = n_{\rm DM}(a)\cdot m(\phi)
\end{equation*}
where $n_{\rm DM}$ is the number density of dark matter.  This is the model we analyze in this paper.

Note that this model captures a wide class of models, including ones where only a fraction $f$ of dark matter is affected by the field $\phi$. In this case, there is a single parameter $f \tilde{c}$ that dictates the cosmology; our results are presented for $f=1$, and can be easily scaled for other values of $f$ in a given model.

For the quintessence potential we choose the same parameterization as in~\cite{Agrawal:2018own} where the slope of the potential is steep in the early universe (with logarithmic slope $\sim 50$) and
becomes shallow in the recent times, where we choose some benchmark examples for recent times with logarithmic derivatives $c = 0.1$ and $0.2$.  For simplicity we consider a sum of two exponential potentials which meet at the point $\phi =\phi_0$ where the dark matter sector begins to get light.  That the transition to shallow slope for the potential should be the same place where the tower of light states appears in the dark sector is a natural expectation. As explained in~\cite{Ooguri:2018wrx}, the tower of states drives the potential itself through the Gibbons-Hawking entropy relation with the dark energy density.  Thus we consider the potential for the scalar to be:
\begin{align}
V(\phi)
&=
B\cdot\exp(-b \phi) +C\cdot\exp(-c \phi)
\end{align}
with $B,C$ chosen so that $B\ \exp(-b \phi_0) = C\ \exp(-c \phi_0)$, and
where $b\sim 50$ and $c=\{0.1,0.2\}$. The choice we make for the logarithmic slope $b$ has no bearing on our result since data cannot distinguish between any value of $b\gtrsim 30$. We assume that in the early universe $\phi$ starts at $\phi <\phi_0$, where it does not couple to dark matter.  This leads to its behavior in early times being set by $V(\phi)$ alone.
The steep part of the potential dominates the behavior which is that of a tracking field with $\Omega_\phi\propto 1/b^2$.  For $b\gtrsim 30$ this is subdominant and does not affect early cosmological evolution.

The energy density today is dominated by the quintessence field with an equation of state close to $-1$. Therefore, the field always transitions through the value $\phi_0$ in the late universe, after which the dark matter--dark energy coupling is active, and the dark matter energy density is converted to dark energy.

It is worth emphasizing the difference between our model and one where the dark matter -- dark energy coupling is always present (e.g.~\cite{Amendola:2003eq,Marsh:2011gr,Pettorino:2012ts,Pettorino:2013oxa}).
The case considered here is in accordance with the behavior expected from string theory, i.e. the coupling is only relevant after $\phi$ has rolled $\mathcal{O}(1)$ in Planck units.
Were the coupling relevant throughout the history of the universe, it would deplete the dark matter density significantly which makes it difficult to circumvent CMB constraints unless $\tilde{c}$ is set to be very small
(see~\cite{Miranda:2017rdk,DiValentino:2017iww} for recent work and references). In this case the coupling is restricted to be $\tilde{c} \lesssim 0.06$.

\section{Results}
\label{sec:results}
Our solution modifies the cosmological evolution of the universe at late times. At early times $(z\gtrsim 15)$, the scalar field contributes a negligible amount to the energy density of the universe.
In this phase of its evolution, the scalar field is evolving on a steep exponential potential. The attractor solution for this case is that the scalar field tracks the background energy density,
\begin{align}
\rho_\phi &= \frac{3(1+w_b)}{b^2} \rho_b
\end{align}
where $\rho_b$ and $w_b$ are the energy density and equation of state of the background cosmology.
In our case $b = 50$, which means that the contribution is too small to affect CMB anisotropies.
Furthermore, in this early phase there is no coupling of the dark matter to the scalar field, corresponding to the fact that the dark matter mass does not depend on the scalar field $\phi$ in this regime. Indeed, the coupling of the scalar field to the dark matter is proportional to $\partial m_{DM}(\phi) / \partial \phi$. Consequently, the early universe cosmology is similar to that of $\Lambda$CDM.

\begin{figure}[t]
    \centering
    \includegraphics[width=0.45\textwidth]{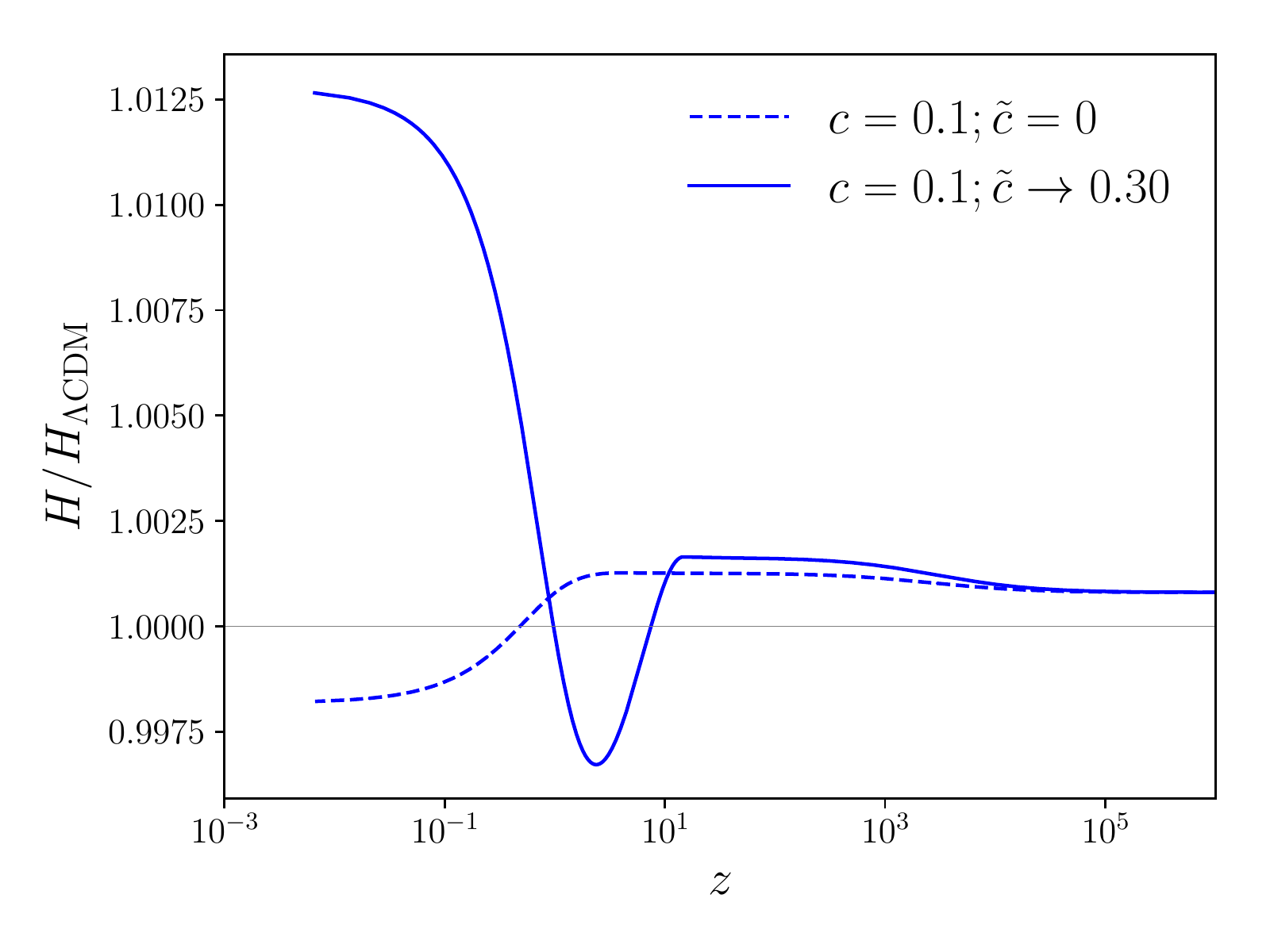}
    \caption{Evolution of the Hubble parameter in our best-fit models with $c=0.1$ relative to that of our best-fit $\Lambda$CDM. Quintessence-only models lead to a smaller value of $H_0$, whereas models with coupled dark matter -- dark energy lead to a larger $H_0$. The coupling becomes relevant at $z\approx 15$.}
    \label{fig:Hplot}
\end{figure}

\begin{figure}[ht]
    \centering
    \includegraphics[width=0.45\textwidth]{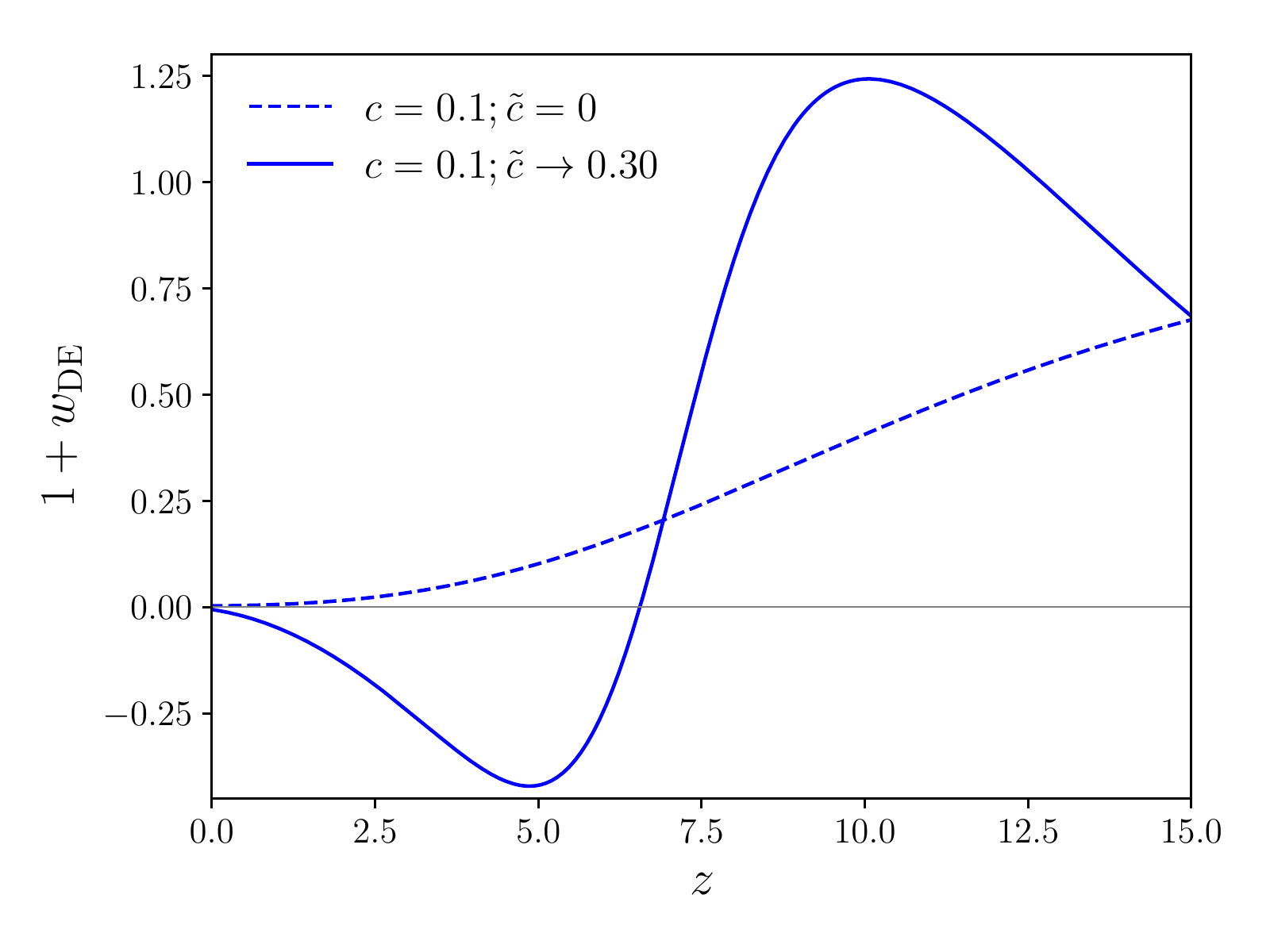}
    \caption{We show the evolution of $w(z)$ in our models. In the presence of coupling of dark matter with the scalar field $(1+w_{\rm DE})$ can be negative as discussed in section~\ref{sec:IIIC}.}
    \label{fig:wplot}
\end{figure}

We show the evolution of $H(z)$ (relative to $\Lambda$CDM) and $w_{DE}(z)$ for our best-fit models
in figures~\ref{fig:Hplot} and~\ref{fig:wplot} respectively.  We provide a precise definition of $w_{DE}(z)$ in equation~\ref{eq:weq} below.
The scalar field coupling to dark matter turns on as the scalar field exits its tracking behavior which typically happens at a redshift $z\sim 15$ in our models. About 10\% of dark matter has faded starting from $z=15$ to the present time at $z=0$. Models which do not couple the scalar field to dark matter tend to have a slightly lower value of $H_0$ than $\Lambda$CDM models.
However, when dark matter and dark energy are coupled, this is no longer true. This coupling depletes dark matter into dark energy $\phi$. As we see from figure~\ref{fig:wplot}, initially, the dark energy equation of state $w_{DE} > 0$, so that the energy density redshifts faster than that in $\Lambda$CDM. This results in a smaller value of $H(z)$ around $z\sim 10$ (see figure~\ref{fig:Hplot}). In order to keep the angular diameter distance to recombination ($D_A$) fixed, the model prefers a larger value of vacuum energy, and hence the value of $H_0$ today is larger.
We see that the dark matter -- dark energy coupling goes in the direction of producing larger values of $H_0$. We will next describe our fitting procedure in detail and present results for the fits.

\begin{table*}[ht]

    \begin{tabular}{
    |
    p{25mm}|
    C{14mm}|
    C{14mm}|
    C{14mm}|
    C{14mm}|
    C{14mm}|
    C{16mm}|}
    \hline
         & $\chi^2$
         &\multicolumn{5}{c|}{$\Delta \chi^2$}
         \\
         \cline{2-7}
         & $\Lambda$CDM
         &\multicolumn{2}{c|}{$c=0.1$}
         &\multicolumn{2}{c|}{$c=0.2$}
         &$c\rightarrow 0.0014$
         \\
         \cline{3-6}
         &
         & $\tilde{c} = 0$ & $\tilde{c} \rightarrow 0.30$
         & $\tilde{c} = 0$ & $\tilde{c} \rightarrow 0.31$
         &
         $\tilde{c}\rightarrow 0.31$
         \\
         \hline
         Planck-high\;$\ell$
         & 2440.12  &  1.12  &  -0.44  &  1.50  &  -0.14  &  -0.74
         \\
         Planck-low\;$\ell$
         & 10496.03  &  0.23  &  -0.13  &  0.38  &  0.028  &  -0.11
         \\
         Planck-lensing
         & 9.48  &  -0.18  &  0.88  &  -0.10  &  0.90  &  1.06
         \\
         BAO
         & 1.80  &  -0.022  &  0.18  &  -0.022  &  0.18  &  0.22
         \\
         Pantheon
         & 1026.89  &  0.066  &  0.086  &  0.13  &  0.046  &  0.19
         \\
        HST
         & 16.56  &  0.72  &  -4.62  &  0.80  &  -4.40  &  -5.30
         \\
        low-$z$ BAO
        & 1.88  &  -0.18  &  0.86  &  -0.16  &  0.86  &  1.00
        \\
         \hline
        Total
         & 13992.77  &  1.77  &  -3.21  &  2.49  &  -2.55  &  -3.71
         \\
         \hline
         Improvement ($\sigma$)
         & -
         & -1.3$\sigma$
         &  1.8$\sigma$
         & -1.6$\sigma$
         &  1.6$\sigma$
         &  1.9$\sigma$
         \\
         \hline
    \end{tabular}
    \caption{Likelihoods for the best-fit point for each of the models considered in this paper. For models where parameters were scanned we indicate the best-fit value by an arrow.}
    \label{tab:fitlikelihoods}
\end{table*}

\begin{table*}
\def\arraystretch{1.5}
\begin{tabular}{|c|c|c|c|c|c|c|}
    \hline
    mean$^{+1\sigma}_{-1\sigma}$
    &\multirow{2}{*}{$\Lambda$CDM}
    &\multicolumn{2}{c|}{$c=0.1$}
    &\multicolumn{2}{c|}{$c=0.2$}
    &$c\rightarrow 0.0014$
    \\
    \cline{3-6}
        bestfit
        &
         & $\tilde c = 0$ & $\tilde c \rightarrow 0.30$
         & $\tilde c = 0$ & $\tilde c \rightarrow 0.31$
         & $\tilde c \rightarrow 0.31$
         \\
         \hline
         \hline
         \multirow{2}{*}{$100\omega_b$}
         & 2.240 $^{+0.015} _{-0.014}$
         & 2.241 $^{+0.014} _{-0.014}$
         & 2.241 $^{+0.014} _{-0.015}$
         & 2.241 $^{+0.014} _{-0.014}$
         & 2.242 $^{+0.014} _{-0.014}$
         & 2.242 $^{+0.015} _{-0.015}$
\\
& 2.237
& 2.237
& 2.237
& 2.241
& 2.238
& 2.236
         \\
         \arrayrulecolor{lightgray} \hline
          \multirow{2}{*}{$\omega_c$}
          & 0.1175  $^{+0.0011} _{-0.0011}$
          & 0.1178  $^{+0.0010} _{-0.0011}$
          & 0.1177  $^{+0.0011} _{-0.0011}$
          & 0.1177  $^{+0.0011} _{-0.0011}$
          & 0.1175  $^{+0.0011} _{-0.0011}$
          & 0.1177  $^{+0.0012} _{-0.0011}$
          \\
          & 0.1177
          & 0.1179
          & 0.1179
          & 0.1177
          & 0.1179
          & 0.1180
         \\
         \hline
         \multirow{2}{*}{$\tau_{\rm reio}$}
         & 0.078 $^{+0.013} _{-0.013}$
         & 0.079 $^{+0.013} _{-0.010}$
         & 0.062 $^{+0.010} _{-0.019}$
         & 0.079 $^{+0.013} _{-0.010}$
         & 0.062 $^{+0.009} _{-0.021}$
         & 0.061 $^{+0.007} _{-0.023}$
         \\
         & 0.077
         & 0.079
         & 0.052
         & 0.081
         & 0.049
         & 0.051
         \\
         \hline
        \multirow{2}{*}{$100\theta_s$}
        & 1.04200  $^{+0.00029} _{-0.00029}$
        & 1.04186  $^{+0.00027} _{-0.00026}$
        & 1.04187  $^{+0.00029} _{-0.00029}$
        & 1.04185  $^{+0.00027} _{-0.00028}$
        & 1.04188  $^{+0.00029} _{-0.00028}$
        & 1.04186  $^{+0.00029} _{-0.00029}$
        \\
        & 1.04199
        & 1.04183
        & 1.04183
        & 1.04185
        & 1.04188
        & 1.04181
         \\
         \hline
        \multirow{2}{*}{$\ln [10^{10}A_s]$}
        & 3.084  $^{+0.024} _{-0.023}$
        & 3.087  $^{+0.023} _{-0.018}$
        & 3.051  $^{+0.024} _{-0.032}$
        & 3.087  $^{+0.024} _{-0.019}$
        & 3.051  $^{+0.023} _{-0.036}$
        & 3.051  $^{+0.023} _{-0.035}$
        \\
        & 3.081
        & 3.085
        & 3.032
        & 3.089
        & 3.025
        & 3.029
         \\
         \hline
        \multirow{2}{*}{$n_s$}
        & 0.9687  $^{+0.0042} _{-0.0043}$
        & 0.9686  $^{+0.0042} _{-0.0041}$
        & 0.9688  $^{+0.0043} _{-0.0042}$
        & 0.9691  $^{+0.0041} _{-0.0041}$
        & 0.9690  $^{+0.0042} _{-0.0042}$
        & 0.9687  $^{+0.0042} _{-0.0042}$
        \\
        & 0.9665
        & 0.9664
        & 0.9659
        & 0.9673
        & 0.9664
        & 0.9658
         \\
          \hline
         \multirow{2}{*}{$\tilde{c}$}
         & -
         & -
         & 0.25 $^{+0.10} _{-0.06}$
         & -
         & 0.25 $^{+0.11} _{-0.06}$
         & 0.25  $^{+0.11} _{-0.06}$\\
         & -
         & -
         & 0.30
         & -
         & 0.31 & 0.31
         \\
         \hline
         \multirow{2}{*}{$r_{\rm drag}$}
         & 147.74  $^{+0.25} _{-0.25}$
         & 147.53  $^{+0.25} _{-0.25}$
         & 147.55  $^{+0.25} _{-0.25}$
         & 147.56  $^{+0.26} _{-0.26}$
         & 147.59  $^{+0.26} _{-0.27}$
         & 147.56  $^{+0.26} _{-0.26}$
         \\
         & 147.72
         & 147.57
         & 147.54
         & 147.56
         & 147.55
         & 147.54
         \\
         \hline
          \multirow{2}{*}{$\sigma_8$}
& 0.8192 $^{+0.0085} _{-0.0088}$ & 0.8161 $^{+0.0085} _{-0.0071}$ & 0.844 $^{+0.016} _{-0.019}$ & 0.8148 $^{+0.0088} _{-0.0072}$ & 0.844 $^{+0.018} _{-0.020}$ & 0.844 $^{+0.019} _{-0.019}$ \\
&0.8180  &0.8145 &0.848 &0.8149 &0.851 &0.852
         \\
         \hline
        \multirow{2}{*}{$H_0$}
        & 68.34 $^{+0.50} _{-0.51}$
        & 68.19 $^{+0.50} _{-0.48}$
        & 69.06 $^{+0.66} _{-0.73}$
        & 68.12 $^{+0.48} _{-0.50}$
        & 68.98 $^{+0.62} _{-0.66}$
        & 68.94 $^{+0.67} _{-0.71}$ \\
&68.25 &68.13 &69.12 &68.11 &69.08 &69.27
         \\
         \arrayrulecolor{black} \hline
    \end{tabular}
    \caption{\label{tab:bestfit} Best fit, mean and $\pm1\sigma$ values for cosmological parameters in $\Lambda$CDM and each of the models considered in this paper.
   }
\end{table*}
\subsection{Fits and Datasets}

We use a modified version of the CMB Boltzmann code \texttt{CLASS} \cite{2011arXiv1104.2932L,2011JCAP...07..034B} with \texttt{MontePython} \cite{Brinckmann:2018cvx,Audren:2012wb} to scan the parameter space of our model. Our model has two free parameters, $c$ and $\tilde{c}$.
We use the following datasets
\begin{itemize}
    \item Planck: We use the temperature and polarization likelihoods from Planck 2015 high-$\ell$~\cite{Aghanim:2015xee}, low-$\ell$ and lensing~\cite{plancklens} likelihoods, including all the nuisance parameters with priors as in~\cite{planck15}.
    \item BAO: We use the BAO measurements from the BOSS-DR12 $f\sigma8$ sample~\cite{Alam:2016hwk}. We also use the low-$z$ BAO measurements from 6dFGS~\cite{BAO1} and Main Galaxy Sample from SDSS~\cite{BAO3}.
    \item Pantheon: We use the Pantheon sample~\cite{Scolnic:2017caz}
    consisting of a total of 1048 SN Ia in a redshift range of $0.01 < z < 2.3$.
    \item SH0ES: We use the recent $H_0$ measurement that adds LMC Cepheids to get a $2\%$ level measurement of $H_0$~\cite{Riess:2019cxk}  as $74.03\pm 1.42$ km/s/Mpc.
    \item We use a flat prior for all $\Lambda$CDM and new physics parameters, where applicable. We restrict $\tau_{\rm reio}>0.038$ in order to be consistent with the observation of the Gunn-Peterson trough in the spectrum of high-redshift quasars~\cite{Aghanim:2018eyx}.
\end{itemize}

\begin{figure}[t]
    \centering
    \includegraphics[width=0.45\textwidth]{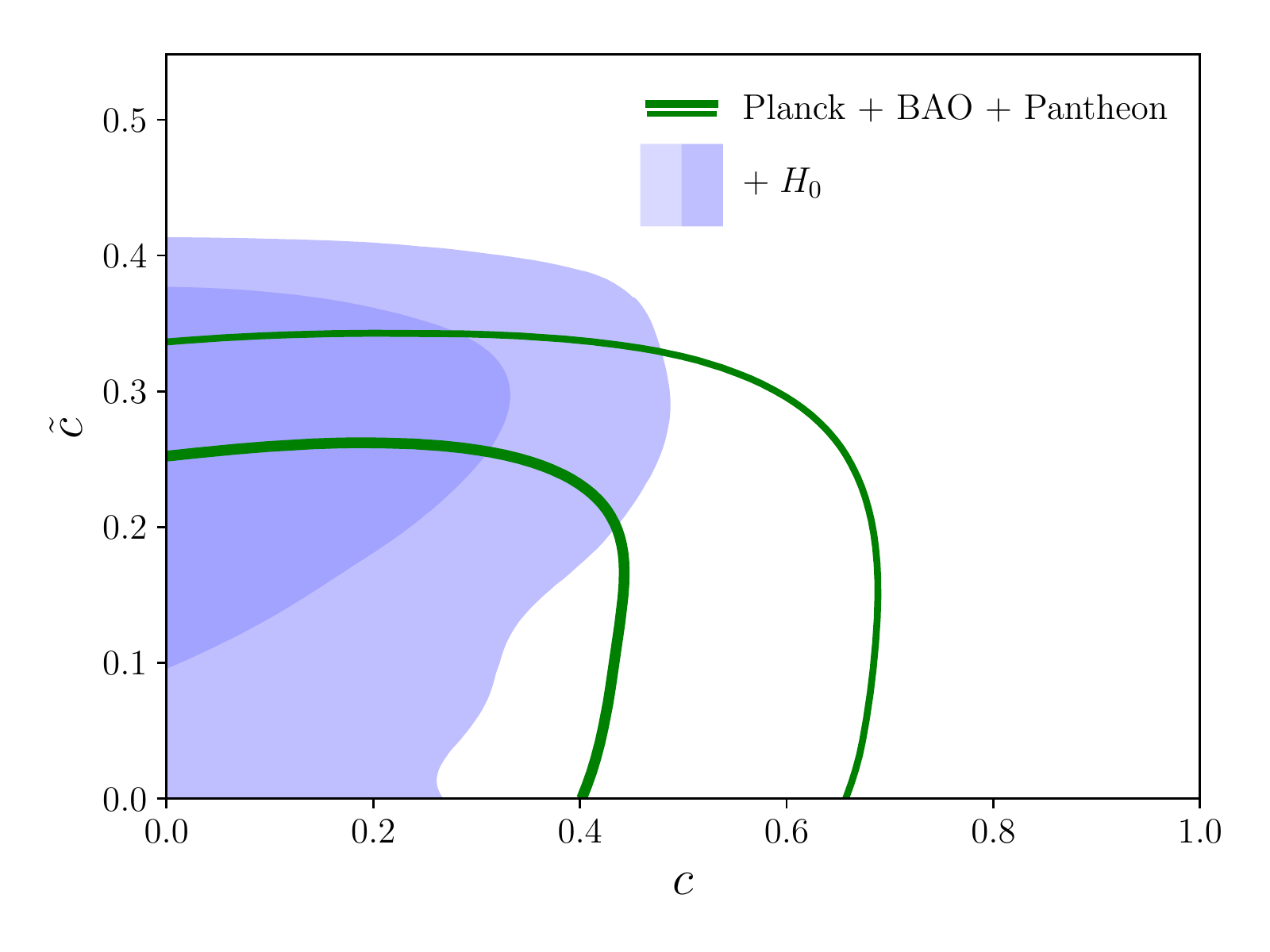}
    \caption{Contours showing 1 and 2$\sigma$ bounds for $c,\tilde{c}$ including HST (shaded blue) and without HST (green).
    We see that including the HST dataset disfavors the $\tilde{c}=0$ point.
    }
    \label{fig:abplot}
\end{figure}

We run parameter scans with two different strategies. Firstly, in order to map out the parameter space, we perform a scan over both our parameters $c$ and $\tilde{c}$. To assess the importance of including the SH0ES likelihood, we also sample from the posterior with the $H_0$ prior from HST omitted. Our posteriors for these runs are shown in figure~\ref{fig:abplot}. The posterior for the parameter $c$ is peaked at $0$, with a $2\sigma$ allowed region of $c < 0.6$ (without HST) and $c<0.4$ (with HST). This is consistent with what was found in a preliminary analysis in~\cite{Agrawal:2018own}. Without the HST data, we also obtain constraints on the parameter $\tilde c < 0.3$. Interestingly, however, we find that the value $\tilde{c}$ is preferred to be non-zero once the $H_0$ data is taken into account, and $\tilde{c}=0$ is disfavored at the 2.8$\sigma$ level.
 This is expected from our discussion earlier and figure~\ref{fig:Hplot}.

Secondly, we fix values of the parameter $c = 0.1,0.2$ and scan over the parameter $\tilde{c}$. This is motivated by the swampland dS conjecture which points to a non-zero $\mathcal{O}(1)$ value for $c$. Consequently, there is a theoretical prior on $c$ away from zero. The results of the corresponding fit are shown in table~\ref{tab:fitlikelihoods}.
We see that for values of $c$ in the range $[0,0.2]$ the fits do not change appreciably.
In each case the presence of $c$ alone  makes the HST likelihood worse.
However, adding $\tilde{c}$ improves this likelihood (even compared to $\Lambda$CDM).
We show the best-fit values as well as the mean and error bars for the cosmological parameters in table~\ref{tab:bestfit}.

\begin{figure}[t]
    \centering
    \includegraphics[width=0.45\textwidth]{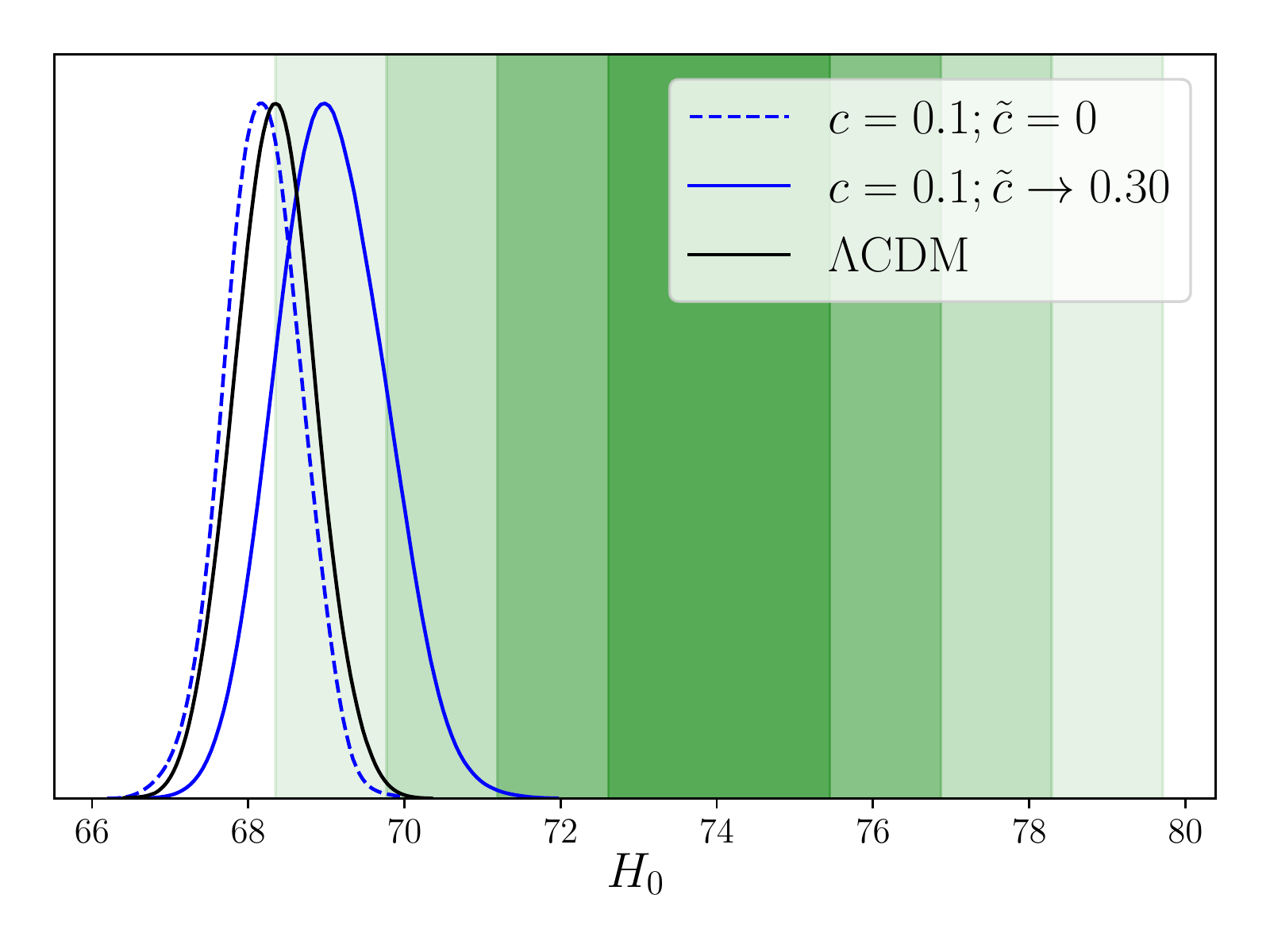}
    \caption{Posteriors for $H_0$ in $\Lambda$CDM, scalar field quintessence and fading dark matter models described in table~\ref{tab:bestfit}. The green shaded regions show the 1, 2, 3 and 4$\sigma$ ranges for the SH0ES measurement~\cite{Riess:2019cxk}.}
    \label{fig:H0distribution}
\end{figure}

In figure~\ref{fig:H0distribution} we show the $H_0$ posterior distribution for the fading dark matter model with $c=0.1$, the $\Lambda$CDM model, and the quintessence model with no coupling to dark matter. Evidently, the model with the coupling to dark matter accommodates a larger value of $H_0$ reducing the tension with the locally measured value from 4.4 to 3.1 $\sigma$.

It is well known that with modifications of late time cosmology, as is the case in our model, the $H_0$ tension cannot be fully resolved~\cite{Bernal:2016gxb,Aylor:2018drw}.  However it is amusing that many solutions offered for fully resolving this tension involve the addition of rolling scalar fields~\cite{Agrawal:2019lmo,Poulin:2018cxd,Lin:2019qug} which can be easily accommodated in our current model (in a way consistent with swampland constraints) by slightly modifying the early-time potential $V(\phi)$. This would allow a short period of energy injection around matter-radiation equality (however, unlike our present model, these models make the $S_8$ tension worse). On the other hand, experiments may converge to a somewhat smaller value e.g. the value observed in~\cite{Freedman:2019jwv} $H_0= 69.8 \pm 1.9\ \mathrm{km/s/Mpc}$, which is perfectly consistent with our best fit model. We have also checked that replacing the late time data for $H_0$ with this lower central value (but assuming a smaller 1\% error) our result for ${\tilde c}\sim 0.3$ still holds. In this sense the parameters of our model have chosen values that produce the maximal shift in $H_0$ allowed through late time cosmology given current data.

\begin{figure*}[p]
    \centering
    \includegraphics[width=0.95\textwidth]{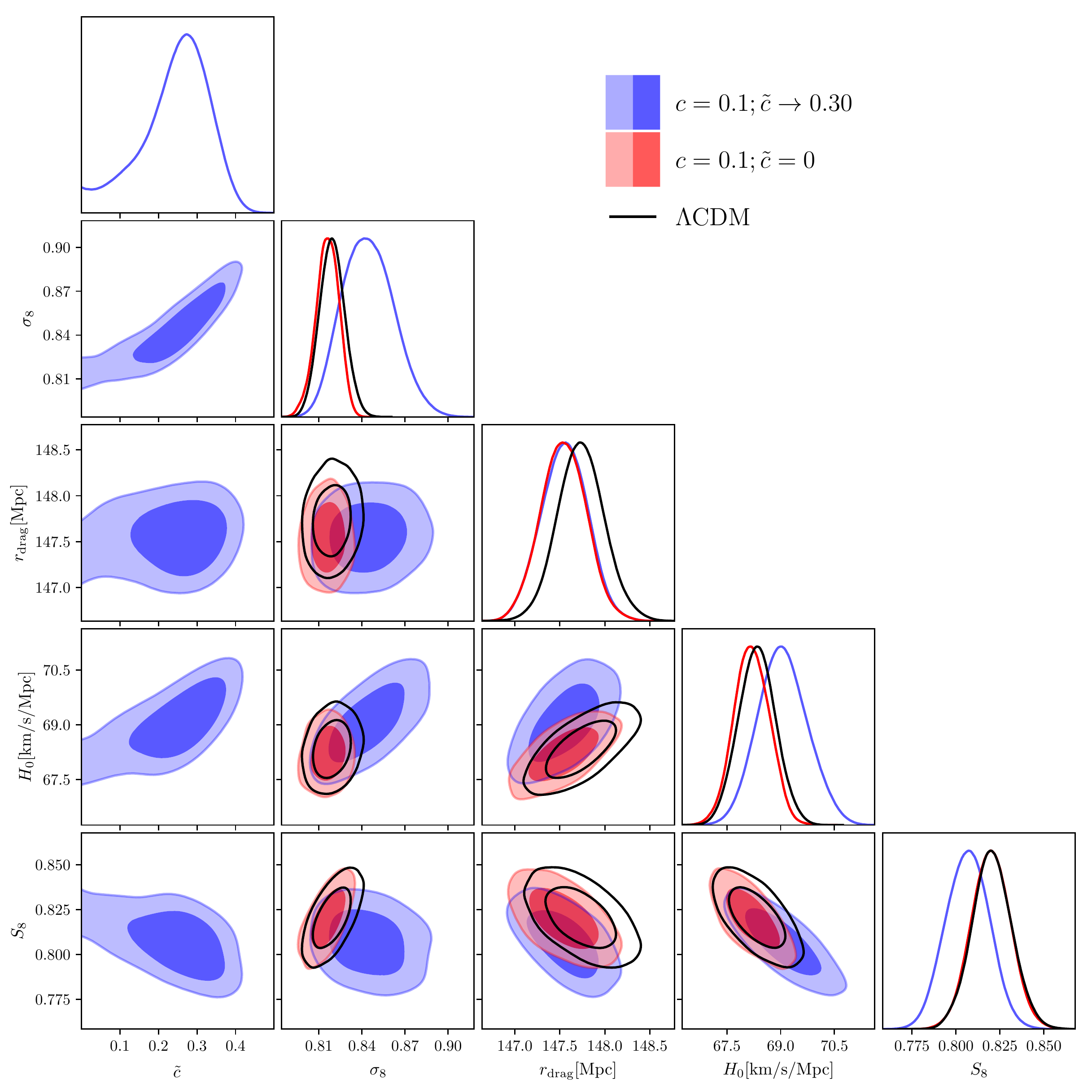}
    \caption{Posterior distributions comparing parameters of the fading dark matter model with $c = 0.1$, scalar field quintessence and $\Lambda$CDM. }
    \label{fig:c0p1distribution}
\end{figure*}

\begin{figure*}[p]
    \centering
    \includegraphics[width=0.95\textwidth]{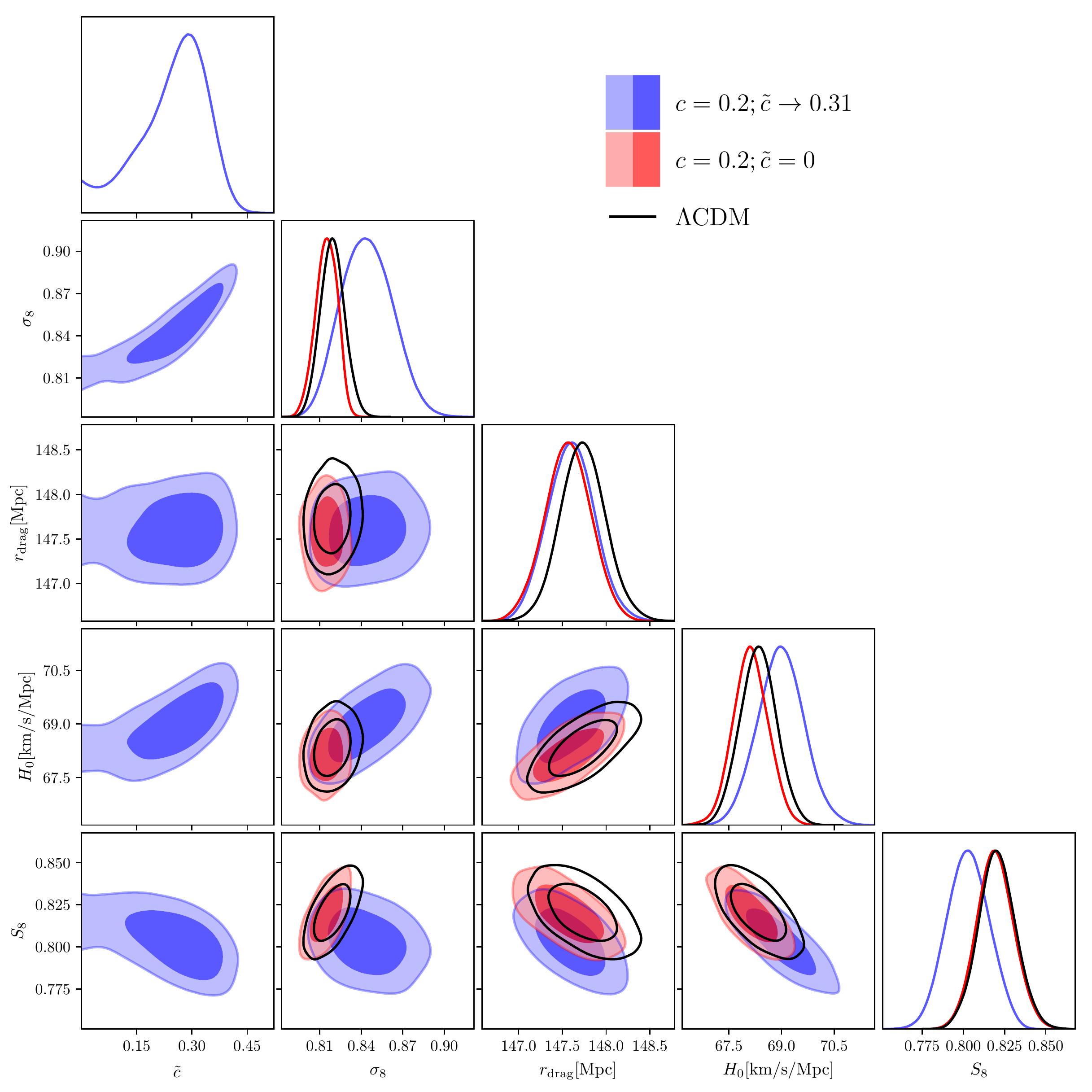}
    \caption{Posterior distributions comparing parameters of the fading dark matter model with $c = 0.2$, scalar field quintessence and $\Lambda$CDM.}
    \label{fig:c0p2distribution}
\end{figure*}

In addition, we show 2D posteriors of the parameters of our fading dark matter model with $c = 0.1$ compared to quintessence and $\Lambda$CDM in figure~\ref{fig:c0p1distribution}. The constraints on $\Lambda$CDM and scalar field quintessence-only model are similar but addition of a coupling to dark matter shifts the $S_8$ and $H_0$ posteriors, as previously mentioned. These effects are more pronounced for larger values of $c$ as shown in figure~\ref{fig:c0p2distribution}. That our model modifies only late time cosmology can be seen from the posterior of $r_{\rm drag}$ which is nearly identical to that of $\Lambda$CDM.

We highlight further phenomenological features of the fading dark matter model below.

\subsection{Matter power spectrum}
The only parameters in early cosmology that are different in our models from $\Lambda$CDM are $\tau_{\rm reio}$ and $A_s$ (such that the well-measured combination $A_s e^{-2\tau_{\rm reio}}$ is preserved). Even though the value of $A_s$ is smaller, our $\sigma_8$ posterior is shifted to larger values than that of $\Lambda$CDM, as a result of late-time modification of the growth rate.

\begin{figure}[t]
    \centering
    \includegraphics[width=0.45\textwidth]{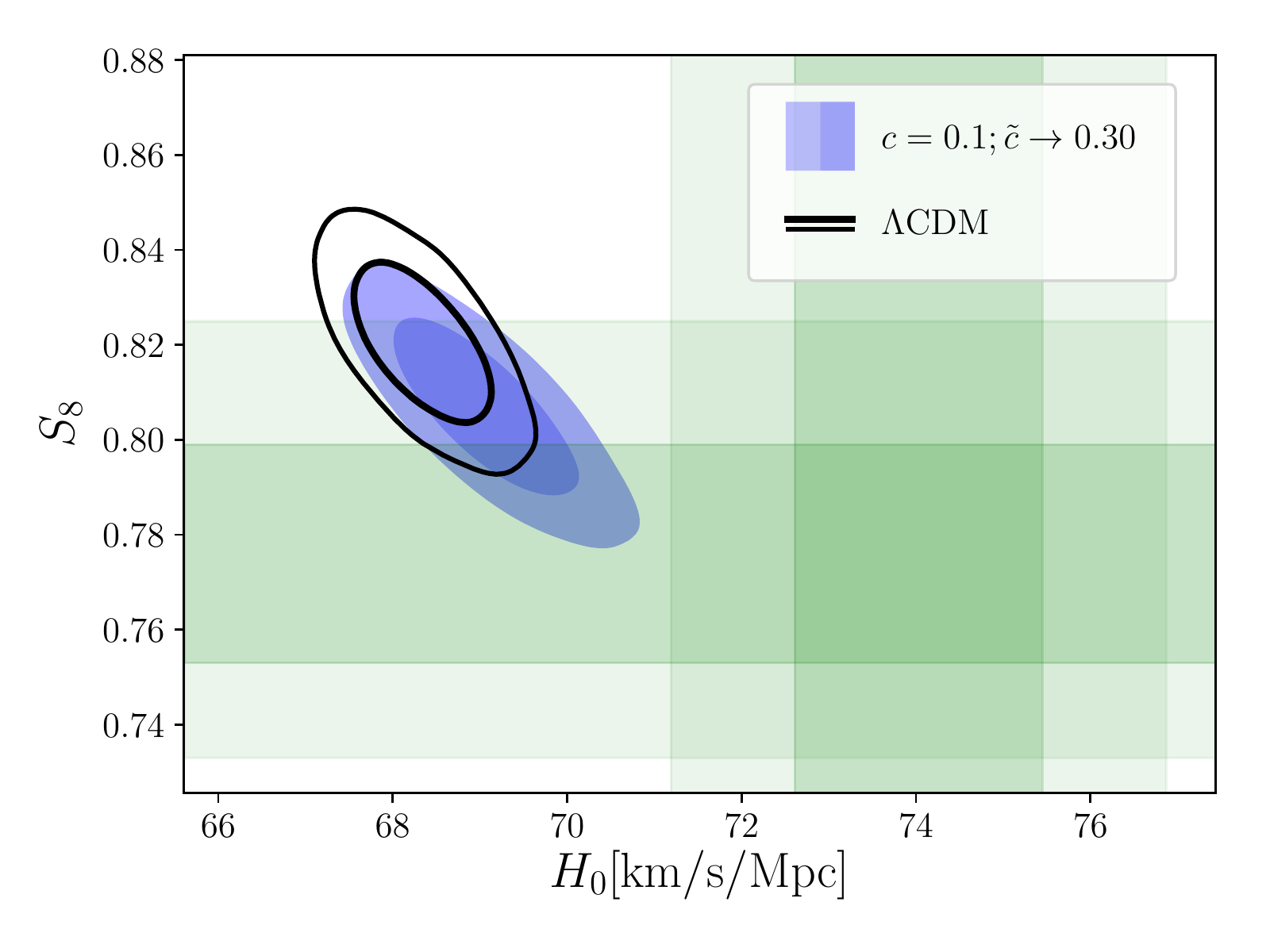}
    \caption{The 2D posterior for $H_0$ and $S_8$ in our model with $c=0.1$ (shaded blue) and $\Lambda$CDM (black) compared to local measurements (green bands). }
    \label{fig:HSplot}
\end{figure}

There is a long-standing tension between measurements of the matter power spectrum, often called the $\sigma_8$ tension, where the value of $\sigma_8$ inferred from the CMB is larger than that measured in weak lensing experiments within $\Lambda$CDM. More precisely, it is the combination $S_8 \equiv \sigma_8 (\Omega_m/0.3)^{0.5}$ that is constrained by lensing measurements. In figure~\ref{fig:HSplot} we show the 2D posterior of $H_0$ and $S_8$ in our fading dark matter model, overlaid with the SH0ES measurement of $H_0$ and the DES measurement of $S_8$~\cite{Abbott:2017wau}. Clearly our model goes in the direction of easing both tensions, but does not resolve them completely. This result is comparable to the results obtained in~\cite{Pandey:2019plg} for a very specific model of decaying dark matter. We note, however, that models of decaying dark matter are tuned in the sense that they require a dark matter lifetime comparable to the age of the universe to have the desired cosmological impact. In addition to its theoretical appeal, fading dark matter makes distinct phenomenological predictions such as a fifth force in the dark sector. It also does not require the presence of dark radiation to which dark matter decays and only affects the late universe whereas generic decaying dark matter can impact early universe cosmology as well.

\subsection{Equation of state for dark energy}
\label{sec:IIIC}
Quintessence models are strongly constrained by supernova measurements~\cite{Scolnic:2017caz,Agrawal:2018own,Heisenberg:2018yae,Raveri:2018ddi} of the equation of state of dark energy, $w_{\rm DE}(z)$. In the presence of a dark matter coupling to scalar fields, there is no invariant definition of $w_{\rm DE}$, since the stress energy tensor of dark energy is not separately conserved.

When fitting to supernova data, we can extract an effective equation of state, $w_{\rm DE}$. This is calculated from the $z$-dependence of $\rho_{\rm DE}$ defined as follows:
\begin{align}
    \rho_{\rm DE}
    \equiv
    \rho_{tot}
    -\rho_{c,0}
    -\rho_{\rm other}
\end{align}
where $\rho_{c,0}$ is the component of dark matter assuming it redshifts like $a^{-3}$ and $\rho_{\rm other}$ includes all other contributions to energy density such as neutrinos, baryons, etc. Then,
\begin{align}
    1+w_{\rm DE}
    &=
    -\frac{1}{3\rho_{\rm DE}}
    \frac{\partial \rho_{\rm DE}}{\partial (\log a)}
    \label{eq:weq}
\end{align}
Note that $1+w_{\rm DE}$ defined this way can in general be negative~\cite{Das:2005yj,vandeBruck:2019vzd}, since the coupled dark matter + dark energy system in our model redshifts slower than a $\Lambda$CDM-like system. This is intriguing since supernova data has a slight preference for a negative value for $1+w_{\rm DE}$ in $w_0w_a$-parametrized equation of state models~\cite{Scolnic:2017caz}.
We show $w_{\rm DE}(z)$ in our best-fit $c = 0.1$ fading dark matter model in figure~\ref{fig:wplot}. Despite this, the Pantheon likelihoods do not have a  preference  for our best fit model relative to the $\Lambda$CDM/quintessence only models.

\subsection{Fifth force in the dark sector}
The coupling of the scalar field $\phi$ with the dark matter induces a
new long-range attractive force exclusively between dark matter
particles. In the non-relativistic limit, this effectively appears as a
modification of Newton's constant for dark matter.

There are a number of constraints on such a modification. Since
$\tilde c \sim 0.3$, the modification is sizeable. Many constraints
rely on
effects on early cosmology~\cite{vandeBruck:2019vzd}. In our model, however, the coupling of the scalar field to dark matter is very small until late times, and consequently the constraints are much weaker. There are also local observables at galactic scales, e.g.~the tidal tails of satellite galaxies of the Milky Way, which are affected by additional forces in the dark sector. The present constraints from the tidal tails~\cite{Kesden:2006vz,Kesden:2006zb} limit the new force to be $\mathcal{O}(10\%)$ (which corresponds to $\tilde c \sim \mathcal{O}(0.1)$)
Therefore, we see that these measurements have the potential to detect this scalar field coupling to the dark matter.

Note that these constraints will be weakened if only a fraction of dark matter is coupled to $\phi$ with a correspondingly larger value of $\tilde{c}$. Even though a part of dark matter is more strongly interacting in this case, most of the gravitational potential felt by baryons is set by ``ordinary" dark matter.

Another late-time constraint is due to studies of galaxy warps~\cite{Desmond:2018euk,Desmond:2018sdy,Desmond:2018kdn}. In the presence of a fifth force in the dark sector, a galaxy-halo system falling in a gravitational potential will have a galactic center that is displaced from the halo center. This displacement causes a U-shaped warping of the stellar disk. Using this effect, a bound is placed on the strength of the fifth force $\Delta G/G\lesssim 10^{-4}$ which would constrain $\tilde{c} \lesssim 10^{-2}$.
However as noted in~\cite{Desmond:2018kdn}\footnote{We would like to thank Harry Desmond for discussions on this.} there are a number of assumptions in reaching this conclusion.  In particular there could be systematics which may suggest that this should not be taken at face value. At the very least this is a powerful way to constrain fifth forces, which is certainly worth further studies.

\clearpage

\subsection{CMB polarization}
\begin{figure}[t]
    \centering
    \includegraphics[width=0.45\textwidth]{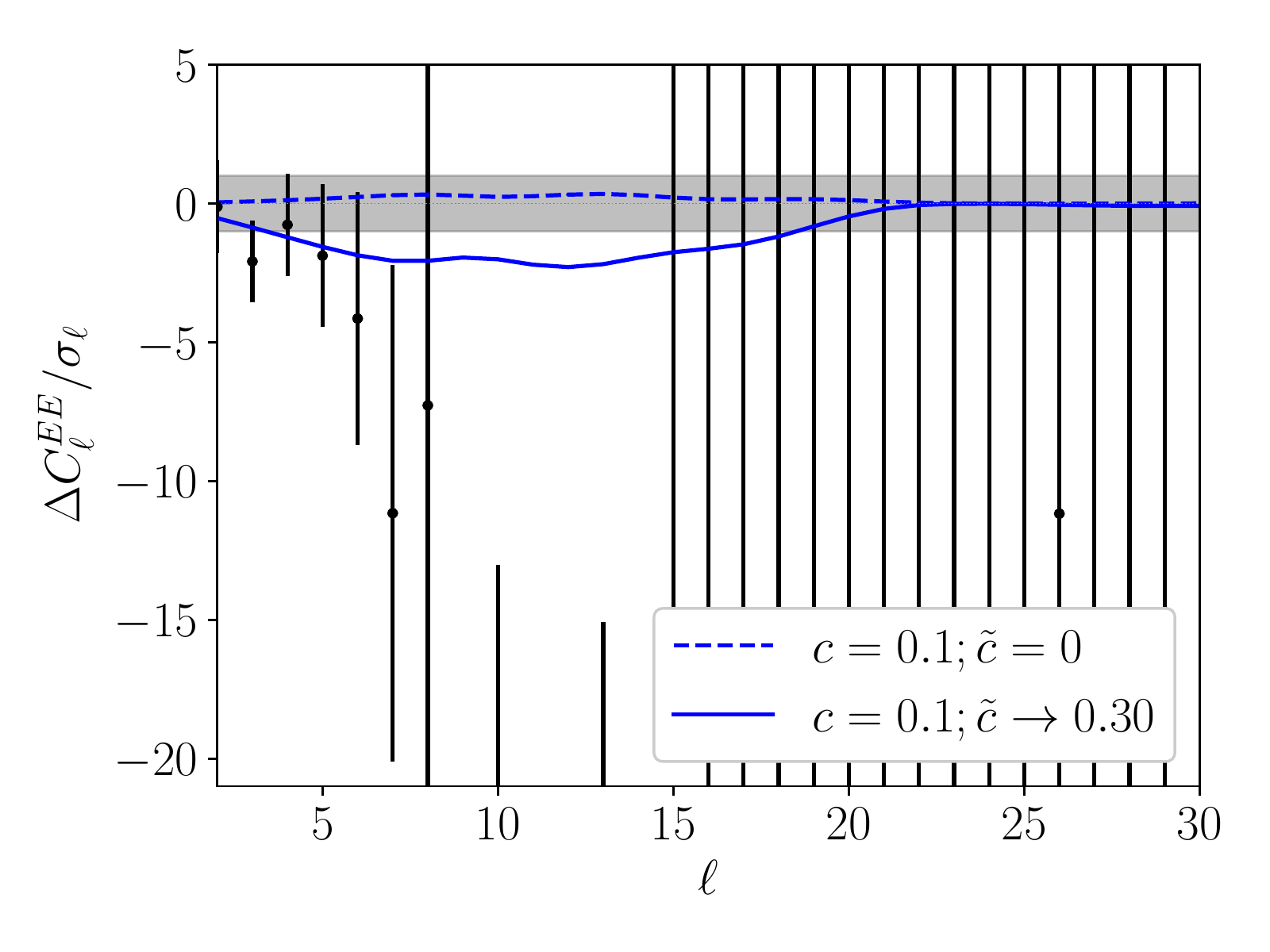}
    \caption{Difference in EE polarization signal between our best-fit fading DM models and $\Lambda$CDM normalized to the cosmic variance per $\ell$ mode. The grey band shows the cosmic variance limit and the data points and errorbars are from {\it Planck} 2015.}
    \label{fig:cmbPol}
\end{figure}

Dark matter fading through coupling to a scalar field alters structure formation (cf. Section III.B). This amplifies the CMB lensing potential and, with all other parameters fixed, worsens the fit to the Planck lensing likelihood. This is remedied in our model by a lowering of the primordial amplitude of fluctuations $A_s$ while holding $A_s \mathrm{e}^{-2\tau}$ fixed. As such, the reionization history is changed which gives an observable effect in CMB polarization power spectra. This effect is shown in figure~\ref{fig:cmbPol} and may be detectable with future CMB polarization experiments such as LiteBIRD~\cite{Matsumura:2013aja}.

\subsection{PREDICTIONS FOR FUTURE}
Similar to the analysis in Ref.~\cite{Agrawal:2018own}, we can extrapolate the scalar field evolution to the future, and predict its impact is on cosmology.

Here we will assume that the scalar field potential remains positive and continues to be of the form chosen, i.e~ $V \sim \exp(-c \phi)$. Since we have already entered the phase of dark energy domination, this term in the potential dominates the future evolution of $\phi$.

On this potential, $\phi$ approaches the fixed point where
\begin{align}
    \Omega_\phi  &= 1,
    \qquad
    d\phi/d (\log a) = c
\end{align}
Therefore, the field travels a distance $1/\tilde{c}$ in  $N=1 /c\tilde c$ number of e-folds. At this point in the moduli space the tower of states start becoming exponentially light, dramatically changing the nature of low-energy physics. For our best fit models, where $c$ lies in the range $[0,0.2]$, we have $\tilde{c}\sim 0.3$. Recall that $c\sim \mathcal{O}(1)$ is theoretically preferred, so we choose $c=0.1$ as a benchmark. This corresponds to $N \simeq 30$ in our fading dark matter model. Intriguingly, this puts the current universe somewhere in the middle of cosmological history from advent of FRW cosmology to its ultimate demise.

\section{Conclusion}
\label{sec:conclusion}
In this note, we considered the implications of the swampland de Sitter and distance conjectures for cosmology. The de Sitter conjecture suggests that the universe today is dominated by an evolving scalar field. The distance conjecture in turn suggests that this evolving scalar field should be responsible for evolution of the mass of a tower of states. Identifying this tower of states with the dark matter, the swampland conjectures motivate a model of evolving dark energy interacting with dark matter.

We studied a simple realization of this setup with the form of the potential and coupling motivated by string theory.
When we include the local $H_0$ measurement, we found the constraints on the parameter $c$ are tighter than those derived in~\cite{Agrawal:2018own}.
These constraints are relaxed (we find at $2\sigma$ that $c < 0.4$), however, if we consider a coupling of the quintessence field with dark matter.
Further, we found that the recent HST measurements of $H_0$ \emph{prefer} a non-zero value of $\tilde{c}\sim 0.3$, a coupling to dark matter, with significance of $2.8\sigma$. This late-time modification improves the fit to data at the $2\sigma$ level compared to $\Lambda$CDM. Taken seriously, this suggests the scale for the $\mathcal{O}(1)$ coefficients in the swampland conjectures.

The fading dark matter model considered here predicts a value of $w(z)$ that
maybe measurable in upcoming dark energy experiments~\cite{Heisenberg:2018yae}. The coupling of dark matter with dark energy produces an additional long-range force between dark matter particles. As mentioned, this can affect the tidal tails of disrupted satellite systems and produce warping of stellar disks in galaxies. The magnitude of the long-range force in our models is expected to produce observable astrophysical signals, and it would be very interesting to look for signs of a new long-range force in the dark sector in astrophysical data.

\section*{Acknowledgments}
We would like to thank
Francis-Yan Cyr-Racine,
Cora Dvorkin,
David Pinner, Lisa Randall, Matt Reece and
Christopher Stubbs
for useful discussions.
 The work of PA is supported by the NSF grants PHY-0855591 and
PHY-1216270.
The research of GO and CV is supported in part by the NSF grant
PHY-1719924 and by a grant from the Simons Foundation (602883, CV).

\bibliography{main}

\end{document}